\documentclass[superscriptaddress,preprint,aps,showpacs]{revtex4}%
\usepackage{amsfonts}
\usepackage{amsmath}
\usepackage{amssymb}
\usepackage{graphicx}
\usepackage{xcolor}%
\providecommand{\U}[1]{\protect\rule{.1in}{.1in}}
\begin{document}
\title{Raman spectrum of 1$T^{\prime}$-WTe$_{2}$ under tensile strain: A
first-principles prediction}
\author{Wei Yang}
\thanks{Author to whom correspondence should be addressed. Electronic mail: yangwei@bupt.edu.cn}
\affiliation{Beijing Key Laboratory of Work Safety Intelligent Monitoring, Beijing
University of Posts and Telecommunications, Beijing 100876, People's Republic
of China}
\author{Zi-Yang Yuan}
\affiliation{Beijing Key Laboratory of Work Safety Intelligent Monitoring, Beijing
University of Posts and Telecommunications, Beijing 100876, People's Republic
of China}
\author{Ying-Qi Luo}
\affiliation{Beijing Key Laboratory of Work Safety Intelligent Monitoring, Beijing
University of Posts and Telecommunications, Beijing 100876, People's Republic
of China}
\author{Yu Yang}
\affiliation{Institute of Applied Physics and Computational Mathematics, Beijing 100088,
People's Republic of China}
\author{Fa-Wei Zheng}
\affiliation{Institute of Applied Physics and Computational Mathematics, Beijing 100088,
People's Republic of China}
\author{Zong-Hai Hu}
\affiliation{Beijing Key Laboratory of Work Safety Intelligent Monitoring, Beijing
University of Posts and Telecommunications, Beijing 100876, People's Republic
of China}
\author{Xiao-Hui Wang}
\affiliation{Beijing Key Laboratory of Work Safety Intelligent Monitoring, Beijing
University of Posts and Telecommunications, Beijing 100876, People's Republic
of China}
\author{Yuan-An Liu}
\affiliation{Beijing Key Laboratory of Work Safety Intelligent Monitoring, Beijing
University of Posts and Telecommunications, Beijing 100876, People's Republic
of China}
\author{Ping Zhang}
\thanks{Author to whom correspondence should be addressed. Electronic mail: zhang\_ping@iapcm.ac.cn}
\affiliation{Institute of Applied Physics and Computational Mathematics, Beijing 100088,
People's Republic of China}
\affiliation{Beijing Computational Science Research Center, Beijing 100084, People's
Republic of China}

\begin{abstract}
Monolayer WTe$_{2}$ attracts rapidly growing interests for its large-gap
quantum spin Hall effect, which enables promising apllicaions in flexible
logic devices. Due to one-dimensional $W$-$W$ chains, 1$T^{\prime}$-WTe$_{2}$
exhibits unique anisotropic structure and promising properties, which can be
modified by simply applying strains. Based on the first-principles
simulations, we show that phonon branch undergoes soft down to negative
frequency at special \textit{q} points under different critical strains,
\textit{i.e.}, $\varepsilon_{a}=11.55\%$ along \textit{a}-axis (with $W$-$W$
chains) direction, $\varepsilon_{b}=7.0\%$ along $b$-axis direction and
$\varepsilon_{ab}=8.44\%$ along biaxial direction. Before each critical
strain, the Raman-shift of $A_{g}^{1}$, $A_{g}^{3}$ and $A_{g}^{4}$\ modes,
corresponding to the main peaks in Raman spectra of 1$T^{\prime}$-WTe$_{2}$,
shows anisotropic response to uniaxial strain but most sensitive to biaxial
strain. Interestingly, we find that the frequency shift of $A_{g}^{3}$ mode
show parabolic characters of strained 1$T^{\prime}$-WTe$_{2}$, then we split
it into two parts and it shows a Raman-shift transition at $\sim$ 5\% strains.
While for the$\mathit{\ }A_{g}^{1}$\ and $A_{g}^{4}$ modes, the frequencies
change linearly. Through careful structure and vibration analysis, we try to
explain these Raman irregularity in strained 1$T^{\prime}$-WTe$_{2}$.

\end{abstract}

\maketitle

\section{ Introduction}

In a robust two-dimensional materials family of transition metal
dichalcogenides (TMDs) \cite{1}, monolayer 1$T^{\prime}$-WTe$_{2}$ has
large-gap quantum spin Hall (QSH) insulator \cite{2,3,4,5,6,7,8,9,10,11},
unconventional spin-torque \cite{12} and gate-tunable superconductivity
\cite{13}, which attracts rapidly growing interests, and enable promising
applications in spintronics, dissipationless transport, and quantum
computations. Among 1$T^{\prime}$-TMDs---MX$_{2}$ with M = (W, Mo) and X =
(Te, Se, S)---theoretically predicted to be a new class of QSH insulators
\cite{2}, WTe$_{2}$ is the only one for which the 1$T^{\prime}$ phase is most
energetically favoured \cite{2} and can be observed under ambient conditions
\cite{1,14}. 1$T^{\prime}$-WTe$_{2}$ with a distorted orthorhombic crystal
structure is distinct anisotropy in contrast to the other monolayer TMDs
crystallized in 1$H$ (trigonal prismatic coordination) or 1$T$ (octahedral
coordination) structures. In plane, the covalently bonded $W$ atoms form a
zigzag $W$-$W$ chain, which makes WTe$_{2}$ structurally one-dimensional and
electronically a semimetal \cite{15}. When cutting 1$T^{\prime}$-WTe$_{2}$
into nanoribbons perpendicular to the $W$-$W$ chains in our previous study
\cite{16}, the electronic band opens a gap, and semimetal transforms to
semiconductor. Other theoretical studies also found that only 1\% tensile
strain along the $W$-$W$ chains can lead to a semimetal to semiconductor or
QSH insulator transition \cite{17,18}. Moreover, the Poisson ration, the
in-plane stiffness and the absorption spectrum of monolayer 1$T^{\prime}%
$-WTe$_{2}$ are strongly dependent on and tunable by tensile strain \cite{17}.

Strain engineering in different orientations can lead to anisotropic
modifications to the structure and properties of monolayer 1$T^{\prime}%
$-WTe$_{2}$, which could play important roles in the application of flexible
logic devices \cite{19}. Theoretical and experimental studies on graphene
\cite{20,21}, monolayer $h$-BN \cite{22,23} and 1$H$-MX$_{2}$ with M = (W, Mo)
and X = (Se, S) \cite{24,25,26,27,28,29} have found that the phonon spectra
are significantly affected by external strains, and their response can be
probed by Raman spectroscopy \cite{30}. In contrast to these 1$H$ structures
with in-plane isotropic, 1$T^{\prime}$-WTe$_{2}$ under stain would exhibit
more interesting anisotropic properties. The Raman spectrum of WTe$_{2}$, from
monolayer, few-layer to bulk as well as alloys
\cite{31,32,33,34,35,36,37,38,39,40}, has been studied to detect the crystal
structure, lattice vibration, number of layers and in-plane anisotropy,
however, the Raman spectrum of strained WTe$_{2}$ has not yet been studied by
either theory or experiment.

Based on the first-principles calculations, we provide a thorough study on the
changes of structures, phonon spectra, and Raman-active modes of monolayer
1$T^{\prime}$-WTe$_{2}$ under uniaxial (parallel or perpendicular to the
$W$-$W$ chains, respectively) and equibiaxial tensile strains. Here, we only
focus on 1$T^{\prime}$-WTe$_{2}$ under tension since a 1$T^{\prime}$-1$H$
phase transition will occur under compression \cite{41}, which may be
experimentally challenging to achieve without incurring any buckling response.
Through our phonon dispersion curve calculations, we find that the acoustic
branches of phonon are the most sensitive to tensile strains, and a soft mode
with negative frequency at different $q$ point under different critical strain
occurs, which indicate the structural instability and signify the ideal
strength of 1$T^{\prime}$-WTe$_{2}$ withstanding larger strain before rupture.
Besides, we also find that the Raman-active modes response quite different to
external strains, and their corresponding frequency changes are anisotopic.
These results indicate that Raman measurements are sufficient to detect the
strain magnitudes and structural stability of 1$T^{\prime}$-WTe$_{2}$.

\section{ Computational method}

Our calculations are based on density functional theory (DFT) and density
functional perturbation theory (DFPT) in pseudopotential plane-wave formalism,
as implemented in the Quantum ESPRESSO package \cite{42}. The atomic positions
and lattice constants for 1$T^{\prime}$-WTe$_{2}$ are optimized by using
Perdew-Burke-Ernzerhof (PBE) \cite{43} functional, and the cutoff energy of 40
Ry and 400 Ry ($1Ry\approx13.6eV$) for the wave functions and the charge
density, respectively. A vacuum spacing $\thicksim13\mathring{A}$\ is used to
prevent the interaction between the WTe$_{2}$ monolayers. A set of 24$\times
1$2$\times1$ $k$-point sampling is used for Brillouin Zone (BZ) integration
over electronic states, and a set of 8$\times$4$\times1$ $q$-point grid is
used for the phonon calculations to obtain the dynamic matrices. The
self-consistent solution of the Kohn-Sham equations is obtained when the total
energy changed by less than 10$^{-8}$ Ry and the Hellmann-Feynman force on
each atom is less than 10$^{-6}$ Ry/bohr ($1bohr\approx0.529177\mathring{A}$).
The lattice constants thus are determined with $a_{0}=3.49\mathring{A}$ and
$b_{0}=6.31\mathring{A}$ and agree with the experimental \cite{14,44} and
theoretica \cite{13,17} values.

\section{ Results and discussion}

The primitive cell of 1$T^{\prime}$-WTe$_{2}$ is indicated by green rectangle,
and its primitive vectors, \ $\overrightarrow{a}$ and $\overrightarrow{b}$,
are indicted by red arrows as shown in Fig. 1(b). Along $a$-axis direction,
the distorted $W$ atoms form 1D zigzag chains indicated by the pink lines. In
the $W$-$W$ chains as show in Fig. 1(c), the stress (corresponding to
$E_{bb}=110.14N/m$) is lager than one in the $b$ axis as shown in Fig. 1(a),
based on our calculation of Young's modulus matrix $E=\left(
\begin{array}
[c]{cc}%
78.59 & 27.68\\
28.43 & 110.14
\end{array}
\right)  N/m$\ which are in agreement with the theoretical \cite{18} and
experimental \cite{34} values. Due to the stronger bond strength in $W$-$W$
chains, one can expect that 1$T^{\prime}$-WTe$_{2}$ could withstand lager
strain along the $a$-axis direction compared to the $b$-axis direction. These
two directions are the subject of our study. A state of uniaxial strain along
$a$-axis and $b$-axis is constructed by applying the nominal strain and ,
respectively, where $a$ and $b$ are the strained lattice constants of
1$T^{\prime}$-WTe$_{2}$.

\begin{figure}[ptb]
\centering
\includegraphics[width= 7.5 cm]{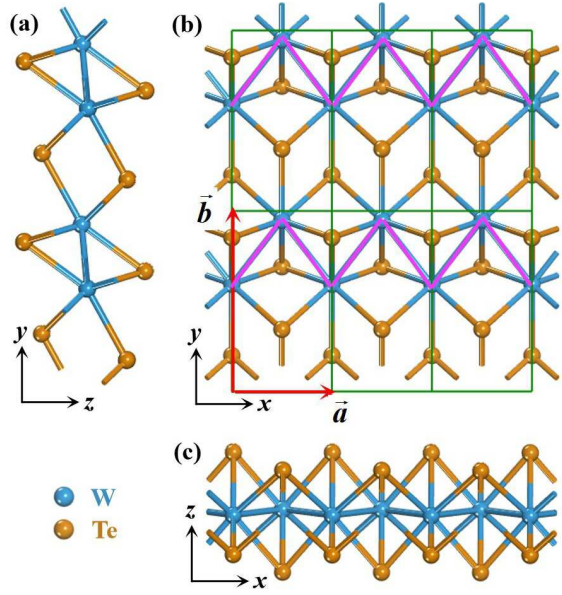}\caption{(color online) Crystalline
structure of monolayer 1$T^{\prime}$-WTe$_{2}$ with (a) side view, (b) top
view and (c) front view. The distorted $W$ atoms form one-dimensional zigzag
chains indicated by the pink lines. The primitive cells (green rectangles) and
the primitive vectors $a$ and $b$ (red arrows) used in the calculations are
shown in (b). }%
\label{fig1}%
\end{figure}

To obtain the phonon and Raman spectra of strained 1$T^{\prime}$-WTe$_{2}$, we
firstly calculate the phonon dispersion curves of intrinsic one, and then
analyse the characteristics of Raman-active modes based on the group theory.
As shown in Fig. 2, there exist 18 phonon branches (3 acoustic and 12 optical
branches) with 6 atoms (2 tungsten and 4 tellurium) in the primitive cell of
1$T^{\prime}$-WTe$_{2}$. Because monolayer WTe$_{2}$ belongs to the No. 11
space group $P21/m$ and the point group $C_{2h}^{2}$, the irreducible
representation of the normal modes at $\Gamma$ point is $6A_{g}+3A_{u}%
+3B_{g}+6B_{u}$, in which only 9 modes are Raman-active among 18 phonon modes
based on the symmetry analysis of calculated atomic displacements. These nine
Rman-active modes as shown in Fig. 2 are $6A_{g}+3B_{g}$, whose frequencies
are shown in Table I with other theoretical \cite{31,32,35,45} and
experimental counterparts \cite{31,32,33,34,35}. Typical Raman spectrum of
1$T^{\prime}$-WTe$_{2}$ exhibits three prominent Raman peaks at $\thicksim
210cm^{-1}$, $160cm^{-1}$, and $130cm^{-1}$ denoted as $A_{g}^{1}$, $A_{g}%
^{3}$ and $A_{g}^{4}$, respectively. Later, we will focus our work exclusively
on these three Raman-active modes.

\begin{figure}[ptb]
\begin{center}
\includegraphics[width= 12.95 cm]{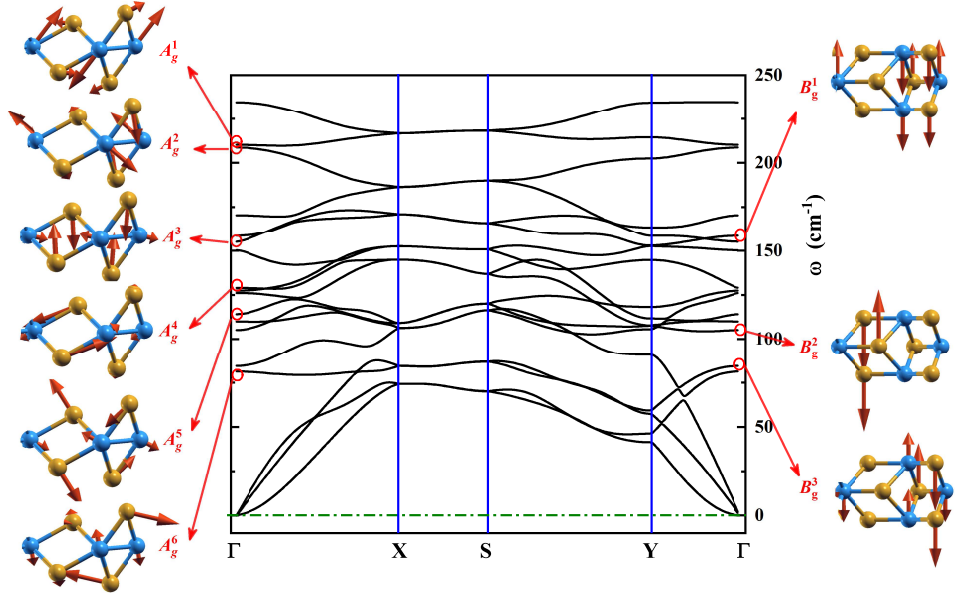}
\end{center}
\caption{(Color online) Calculated phonon dispersion relations and atomic
vibration displacements with their corresponding irreducible representations
for nine typical Raman-active modes (marked by the red circles at $\Gamma$) in
monolayer 1$T^{\prime}$-WTe$_{2}$. Six $A_{g}$ modes with atoms vibrating in
$bc$ plane are listed on the left panel, and three $B_{g}$ modes with atoms
vibrating in $ab$ plane are listed on the right panel. The $q$ point labels
$\Gamma,X,S,Y$ correspond to (0,0), (1/2,0), (1/2,1/2), (0,1/2), respectively,
in fraction of the reciprocal lattice vectors. }%
\label{fig2}%
\end{figure}

\begin{table}[pbh]
\caption{The phonon frequencies of nine Raman-active modes at $\Gamma$ point
for 1$T^{\prime}$-WTe$_{2}$ from Raman experiments and first-principles
calculations. The unit of frequency has been converted to $cm^{-1}$ for direct
comparison with the Raman measurements.}%
\begin{tabular}
[c]{ccccccccccc}\hline
symmetry & Exp.[31] & Exp.[32] & Exp.[33] & Exp.[34] & Exp.[35] & Cal.[31] &
Cal.[32] & Cal.[35] & Cal.[45] & Cal.(this work)\\\hline
$A_{g}^{1}$ & 213.2 & 217 & 212 & 216 &  & 222.6 & 219 & 221.9 & 208.9 &
210.3\\
$A_{g}^{2}$ &  &  &  &  & 216.1 & 220.3 &  & 220.5 & 208.4 & 208.7\\
$B_{g}^{1}$ &  &  &  &  &  & 185.7 &  & 173.7 & 161.3 & 158.5\\
$A_{g}^{3}$ & 161.9 & 164 & 165 & 164 & 164.3 & 164.5 & 168 & 165.9 & 156.5 &
155.1\\
$A_{g}^{4}$ & 132.8 &  &  &  & 135.9 & 137.2 & 135 & 136.9 & 129.9 & 128.9\\
$A_{g}^{5}$ &  &  &  &  & 119.7 & 118.5 & 119 & 119.7 & 114.4 & 114.1\\
$B_{g}^{2}$ &  &  &  &  & 109.4 & 108.5 &  & 110.4 & 105.8 & 105.1\\
$B_{g}^{3}$ &  &  &  & 86 & 88.9 & 93.2 & 89 & 92.5 & 87.2 & 84.5\\
$A_{g}^{6}$ &  &  & 81 &  & 85.7 & 81.4 &  & 84.6 & 80.0 & 81.4\\\hline
\end{tabular}
\end{table}

Owing to strongly anisotropic mechanical properties of 1$T^{\prime}$-WTe$_{2}%
$, we expect that the three Raman-active modes exhibit diverse characteristics
for WTe$_{2}$ under different-direction tensions. Therefore, we then calculate
the phonon dispersion relations of strained monolayer WTe$_{2}$ along
$a$-axis, $b$-axis as well as $ab$-biaxis direction, respectively.
Interestingly, as red curves shown in Fig. 3, the phonon branches become soft
and their frequencies become negative at special $q$ points under different
critical strains. For example, when the tensile strain along $a$-axis
direction approaches 11.55\% ($\varepsilon_{a}=11.55\%$), one of the acoustic
modes of 1$T^{\prime}$-WTe$_{2}$ becomes imaginary near $S$ point indicating
structural instability, which consists with Torum's ($\thicksim11\%$)
\cite{17} and Xiang's ($\thicksim12.5\%$) \cite{18} phonon spectra. In
particularly, applying strains along b-axis direction, Torun \textit{et al}
\cite{17} found that the strain-stress curve reaches a maximum at the critical
strain of 15\% (but without supporting phonon dispersion curves), and Xiang
\textit{et al} \cite{18} found that a phonon branch becomes unstable near the
$\Gamma$ point (not pronounced) at the critical strain $\varepsilon_{b}=12\%$.
However, different from these two works, our study of tensile strain along
$b$-axis direction shows that the ideal strength of monolayer 1$T^{\prime}%
$-WTe$_{2}$ is 7\% with the soft mode occurring at $Y$ point exactly, which is
smaller than that of WTe$_{2}$ along $a$-axis direction with $\varepsilon
_{a}=11.55\%$. Such anisotropic phonons might be attributed to the 1D
structure formed by the $W$-$W$ chains. Meanwhile, we also calculate the
phonon dispersion curves of 1$T^{\prime}$-WTe$_{2}$ under uniform biaxial
strain, as shown in Fig. 3(c), a soft mode occurs near $X$ point at critical
strain $\varepsilon_{ab}=8.44\%$\ indicating a possible phase transition with
the 1$T^{\prime}$-WTe$_{2}$ fracture.

\begin{figure}[ptb]
\begin{center}
\includegraphics[width= 17.5 cm]{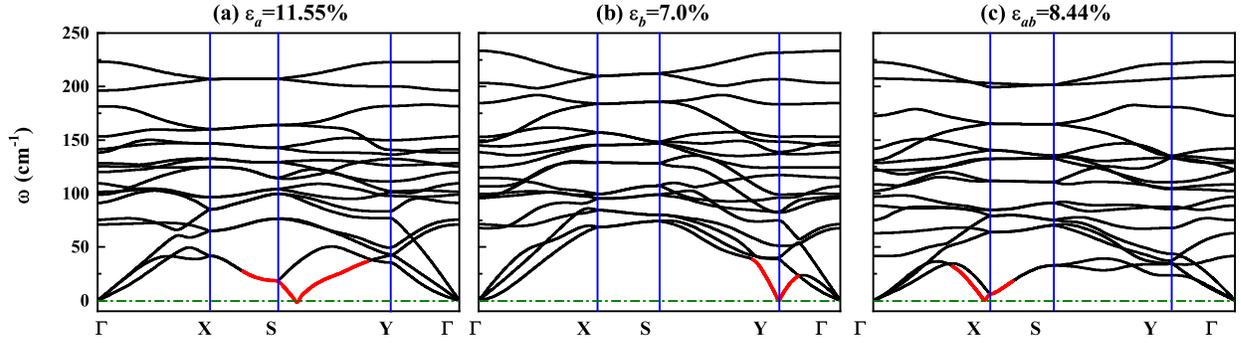}
\end{center}
\caption{(Color online) Calculated phonon dispersion relations for monolayer
1$T^{\prime}$-WTe$_{2}$ at different critical strains (a) $\varepsilon
_{a}=11.55\%$ and (b) $\varepsilon_{b}=7.0\%$ uniaxial tension along $a$ and
$b$ direction, respectively, and (c) $\varepsilon_{ab}=8.44\%$ equibiaxial
tension. The red phonon dispersion curves show negative-frequency modes (named
soft modes) at special $q$ points, indicating the structural instability. }%
\label{fig3}%
\end{figure}

Finally, we analyse the response of Raman-active modes ($A_{g}^{1}$,
$A_{g}^{3}$ and $A_{g}^{4}$) on the three direction strains before
1$T^{\prime}$-WTe$_{2}$ rupture. Due to $A_{g}$ modes containing
double-rotation symmetry-axis operation $C_{2}(z)$ based on the character
table for point group $C_{2h}$, they belong to tangential shear modes with
opposite direction vibrations of $W$-$W$ or $Te$-$T$e pairs. Moreover, the
vibration displacements of $A_{g}$ modes (see Fig. 2) are in the $bc$ plane of
1$T^{\prime}$-WTe$_{2}$, perpendicular to the $a$-axis direction, therefore,
one can expect that the effects of $a$-axis strain on the $A_{g}$ modes would
be weaker than that of $b$-axis strain. As a result, $A_{g}^{4}$ mode stays
relatively unchanged when 1$T^{\prime}$-WTe$_{2}$ under uniaxial strain along
$a$-axis direction, with small slope value of $-0.29cm^{-1}/\%$, as the red
line shown in Fig. 4(c). In order to make the frequency/strain ratio of three
modes comparable, we put them in one figure with the same scale of the
frequency and strain graduations, respectively. And in Fig. 4, we do find that
the frequency shifts in three modes with $b$-axis strain are all much larger
than that with $a$-axis strain, indicating a clear Raman modes anisotropy.
This anisotropic Raman-strain response can facilitate the determination of the
cystallographic orientation in 1$T^{\prime}$-WTe$_{2}$. Moreover, when
applying biaxial strain to 1$T^{\prime}$-WTe$_{2}$, the frequencies of three
Rman modes also red-shift, but exhibit most sensitive to biaxial strain
compared to uniaxial strains, with corresponding largest slope values. To
explain this, we plot Figure 5. In Fig.5, one can find that, although the bond
angle of $W$-$W$ chain is almost unchanged when 1$T^{\prime}$-WTe$_{2}$ under
equibiaxial strain, the $W$-$W$ bond length, the vertical distances in two
pairs of $Te$ atoms ($c_{1}$ and $c_{2}$, respectively) are all dramaticly
changed compared to the unixal strains. This structure modification under
biaxial strain, in turn, leads to a considerable change in the Raman-active modes.

\begin{figure}[ptb]
\begin{center}
\includegraphics[width= 7 cm]{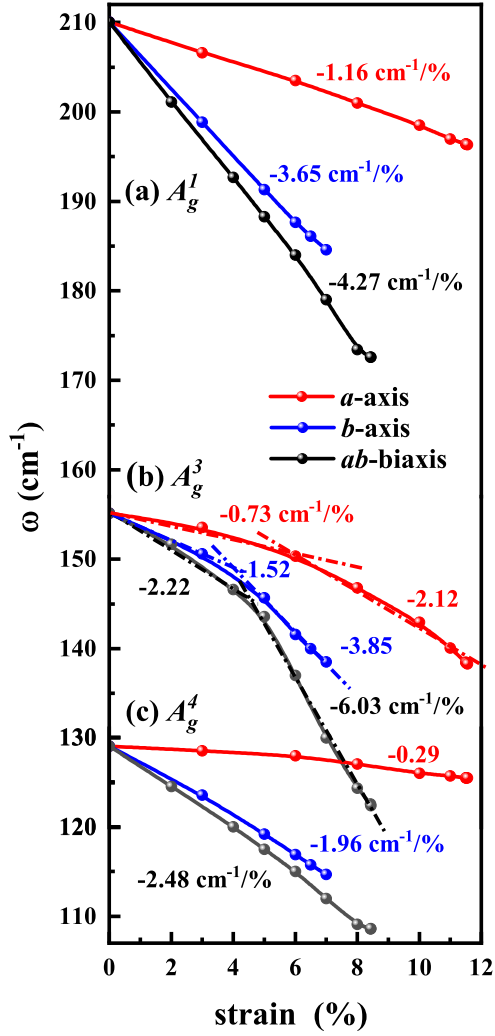}
\end{center}
\caption{(Color online) The phonon frequencies of (a) $A_{g}^{1}$, (b)
$A_{g}^{3}$, and (c) $A_{g}^{4}$ modes (corresponding to three Raman peaks
experimentally) versus strains along $a$-axis (red), $b$-axis (blue), and
$ab$-biaxis (black) directions, respectively. The solid line are linear fits
to the calculated frequencies (solid symbols) and the corresponding slope
values are shown.}%
\label{fig4}%
\end{figure}

\begin{figure}[ptb]
\begin{center}
\includegraphics[width= 14.0 cm]{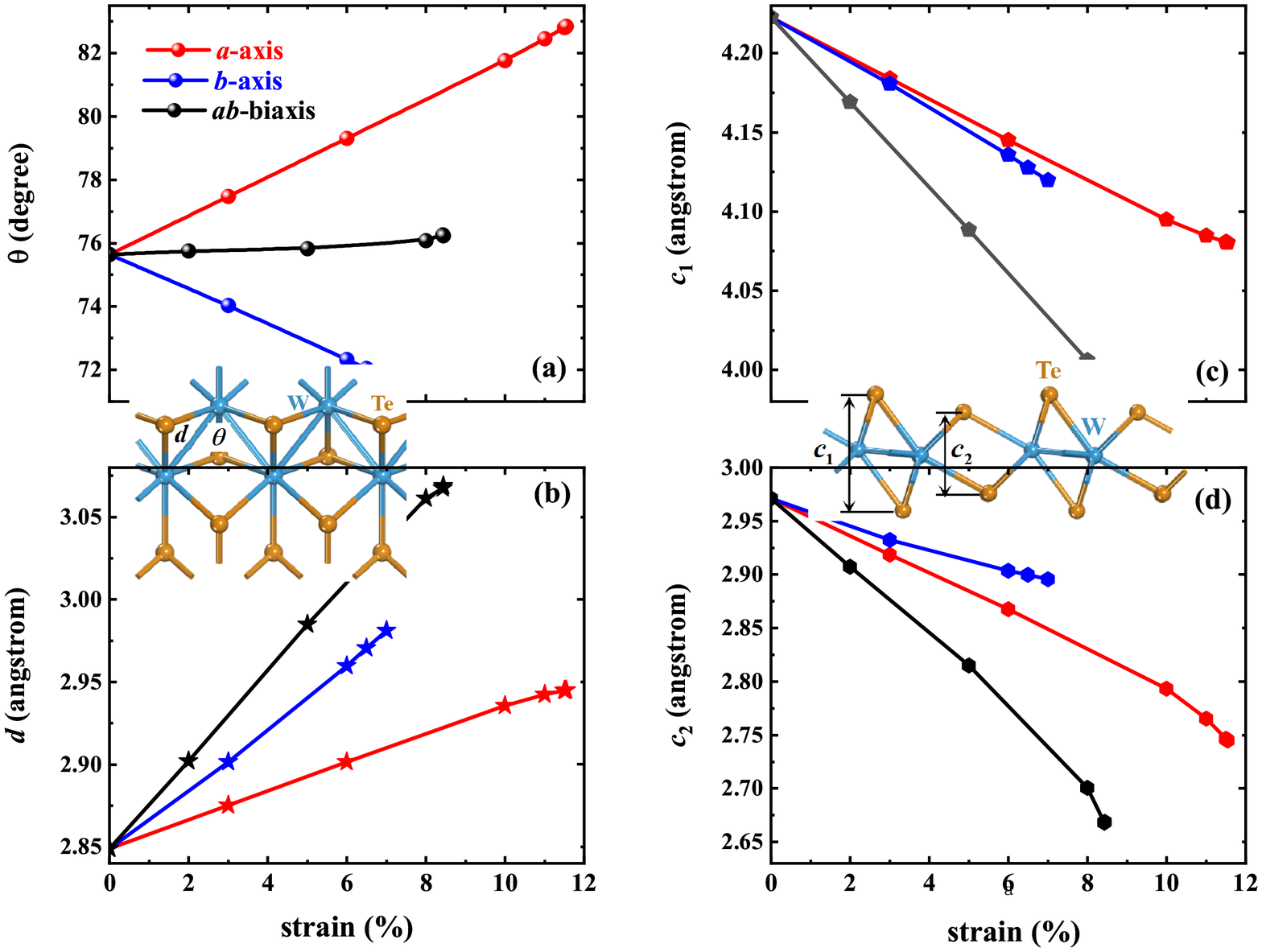}
\end{center}
\caption{(Color online) The calculated (a) $W$-$W$-$W$ bond angle $\theta$,
(b) $W$-$W$ bond length $d$, (c) and (d) vertical distance of two pairs of Te
atoms in $c$ axis, i.e., $c_{1}$ and $c_{2}$, respectively, as functions of
$a$-aixs strain (red lines), $b$-axis strain (blue lines) and biaxial strain
(black lines). The inset crystal configurations are used to depict these
calculated variables. }%
\label{fig5}%
\end{figure}

For $A_{g}^{1}$\ and $A_{g}^{4}$ modes, their frequencies decrease almost
linearly with applied strains as shown in Fig. 4(a) and (c), respectively,
therefore, the strain in 1$T^{\prime}$-WTe$_{2}$ can be quantified by
$A_{g}^{1}$ or $A_{g}^{4}$ mode in Raman spectra. However, for $A_{g}^{3}%
$\ mode, the frequency shift undergoes a sharp turn at about 5\% for both
uniaxial and biaxial strains, as shown in Fig. 4(b), for example, the
calculated biaxial-strain coefficient for $A_{g}^{3}$\ mode is down from
$-2.22cm^{-1}/\%$ to $-6.03cm^{-1}/\%$. This redshift transition might be
result from that the lattice symmetry is maintained before 5\% strains and
then broken after 5\% strains applied. Because the $A_{g}^{3}$ mode consists
of both in-plane and out-of-plane motions as shown in Fig. 2, with regard to
the relative motions in pairs of $W$ atoms (in-plane) and pairs of $Te$ atoms
(out-of-plane), the parallelogram without $W$-$W$ chains could maintain its
shape, but the parallelogram with $W$-$W$ chains might be flattened or
prolongated. Moreover, the amplitudes of $Te$ atoms are much larger than that
of $W$ atoms, especially external strains strengthen the motions of $Te$ atoms
but weaken the vibrations of $W$ atoms, which would prompt the symmetry broken
at last. So, for this particular $A_{g}^{3}$ transition to external strains
and $A_{g}^{3}$\ as the most significant peak with the largest intensity in
Raman spectra, we hope this phenomena can be tested by future Raman
experiments of strained 1$T^{\prime}$-WTe$_{2}$.

\section{Conclusions}

In conclusion, by using DFT and DFPT simulations, we have systematically
obtained the Raman spectrum of monolayer 1$T^{\prime}$-WTe$_{2}$ under tensile
strains. Our results demonstrate that $A_{g}^{1}$, $A_{g}^{3}$ and $A_{g}^{4}$
modes, corresponding to the main peaks in Raman spectra of 1$T^{\prime}%
$-WTe$_{2}$, are more sensitive to uniaxial strain along $b$-axis direction
than that along $a$-axis (containing $W$-$W$ chains) direction, exhibiting
anisotropic Raman-strain response. In particular, biaxial strain induces the
largest red-shift of the Raman modes compared to the uniaxial strains. Through
careful structure configuration analysis, we reveal that the irregularity
comes from the dramatic change of the $W$-$W$ bond length as well as the
vertical distances in two pairs of Te atoms under biaxial strain. More
interesting, $A_{g}^{3}$ mode undergoes a Raman-shift transition at about 5\%
for both uniaxial and biaxial strains, different from the $A_{g}^{1}$ and
$A_{g}^{4}$ modes whose frequencies show linear dependence on tensile strains.
The underlying symmetry reason is revealed to be that the competition and
cooperation of the relative motions between $W$-atom pairs and $Te$-atom pairs
modify the shape of 1$T^{\prime}$-WTe$_{2}$. The revealed Raman modes
responses to tensile strain are useful for tracing or detecting operations on
1$T^{\prime}$-WTe$_{2}$ based flexible devices.

\begin{acknowledgments}
This work was supported by the National Key R\&D Program of China under Grant
Nos. 2017YFB0403602 and 2016YFB0400603, the National Natural Science
Foundation of China under Grant No. 61605014, and the China PFCAEP under Grant
No. YZJJLX2016010.
\end{acknowledgments}

\end{document}